\documentclass[prl,twocolumn,showpacs,preprintnumbers,amsmath,amssymb]{revtex4}
\usepackage{graphicx}
\usepackage{dcolumn}
\usepackage{bm}
\begin{document}

\title{Two-Impurity Anderson-Kondo Model:\\
 first order Impurity-correlation}
\author{J. Simonin}
\affiliation{Centro At\'{o}mico Bariloche and Instituto Balseiro, \\
8400 S.C. de Bariloche, R\'{i}o Negro,  Argentina}
\date{september 2005}

\begin{abstract}
We analyze the consequences of a first-order (in the effective Kondo coupling) inter-impurity
correlation present in the Two-Impurity Anderson Hamiltonian. This correlation generates a pair of Kondo Doublet states which are responsible for the first (higher energy) Kondo stage of these systems. Applied to systems of quantum dots in a semiconductor heterolayer, these Doublets predict a  mainly ferromagnetic  dot-dot spin correlation which is different than the expected from the second-order RKKY interaction. In order to effectively screen both dots, a parallel alignment of the dot-spins is favored by the interaction.
\end{abstract}
\pacs{73.23.-b, 72.15.Qm, 73.63.Kv, 72.10.Fk}
\maketitle

The Anderson Impurity Hamiltonian describes the interaction of
a localized electron state (with a high internal Coulomb interaction) with
a set of uncorrelated-extended orbitals. Primary used to study the behavior of magnetic impurities in a metallic host\cite{coq,varma,wilson}, it also applies to systems of quantum dots (QD) connected to/through a low-dimensional electron gas (LDEG), as has been certified by the observation of the Kondo effect in a single quantum dot, a pioneer work of Goldhaber-Gordon and coworkers\cite{kondoqd}. QD states with a well-defined number of electrons tend to be stabilized by the Coulomb blockade effect. When an odd number of electrons is stable in the dot and its total spin is $1/2$, the hybridization coupling with the LDEG gives rise to the usual Kondo effect.

A logical step in the development of nanoscopic electronics is to considerer systems with several QD\cite{exptk,expli,craig,pascal}, in fact arranges of QD has been proposed with the localized QD-spin ($\bf{S_i}$) state playing the role of a Quantum-bit. For such devices to be useful, a detailed knowledge of the behavior of the QD-spin correlations $\langle \bf{S_i.S_j} \rangle$ is needed. This correlation is expected to be contained in the two-impurity Anderson model. 

The two-impurity Anderson Hamiltonian has been the subject of many theoretical studies, ranging from perturbation theory and narrow band approximation \cite{coq, rkky, cesar, robert} to numerical renormalization group analysis \cite{jay,jones,jay2,wilkins,Kevin,silva} and conformal field theory \cite{pschlot}. An other theoretical approach to Anderson Impurity systems has been the use of variational wave functions (VWF) \cite{varma,saso,sasokato,lucio}. In Ref. \cite{lucio} a set of VWF equations that describes the lower energy state of the system has been numerically analyzed, these equations were analytically solved in Ref.\cite{jsdoublet,jsfull}. These VWF are particularly useful to analyze internal correlations between the components of the system because they are constructed following such correlations. 

In what follows we analyze the consequences of a very simple inter-dot correlation that generates a pair of correlated Kondo Doublets states. Surprisingly, these Kondo Doublets generates a mainly ferromagnetic QD-spin correlation without resort to the second-order RKKY interaction. 

The Anderson Hamiltonian for magnetic impurities (or QD) connected to a metallic electron gas (a LDEG) is the sum of the band $H_b(e_k)$ and effective QD $H_d(-E_d,U)$ Hamiltonians plus the hybridization term $H_V(V)$
\begin{eqnarray}\label{hamil}
H=\sum_{k \sigma} e_k \ c^\dag_{k\sigma} c_{k\sigma} +
\textbf{v} \sum_{j k \sigma}(e^{i\;\bf{k.r_j}} \ \
d^\dag_{j\sigma} c_{k\sigma}+ h.c. )\nonumber \\
 - E_d \sum_{j \sigma} d^\dag_{j\sigma} d_{j\sigma} +
 U \sum_j d^\dag_{j\uparrow} d_{j\uparrow}d^\dag_{j\downarrow}d_{j\downarrow}\ , \ \ \ \ \ \ \ \ 
\end{eqnarray}
where the fermion operator $c_{k\sigma}$($d_{j\sigma}$) act on the conduction
band $k$-state (on the QD state situated at $\bf{r_j}$). Single state energies $e_k, -E_d$ are refereed to the Fermi energy ($E_F$),  $\textbf{v}=V/\sqrt{N}$ is the $d-c$ hybridization $V$ divided by square root of the number of cells in the metal ($N$). In the Kondo limit, the case analyzed in this paper, the QD level is well below the Fermi energy ($-E_d \ll 0$), and they can not be doubly occupied due to the strong Coulomb repulsion in them ($U \rightarrow \infty$). We renormalize the
vacuum (denoted by $|F\rangle$) to be the conduction band filled
up to the Fermi energy and we make an electron-hole transformation
for band states below the Fermi level: $b^\dag_{k\sigma}\equiv
c_{-k,-\sigma}$ for $|k|\leq k_F$; with the above renormalization
the hole excitation energies are explicitly positive.

We consider here the two QD case, one QD at $-
\textbf{r}/2$ and the other at $\textbf{r}/2$, over the $x$-axis.
We use a ``ket" notation for the QD population, the first symbol
indicates the status of the left QD and the second one the status
of the QD on the right, e.g. $|\!\! \downarrow \uparrow \rangle
\equiv d^\dag_{L\downarrow} d^\dag_{R\uparrow} |F\rangle$, $\ | 0
\! \uparrow \rangle \equiv d^\dag_{R\uparrow}|F\rangle$,
 $\ | 0 0 \rangle \equiv|F\rangle$, etc..

The two relevant Hamiltonian parameters in the regime detailed above result to be the effective Kondo coupling $J_n=n_o V^2/E_d$ ($n_o$ being the density of band states
at the Fermi level) and the inter-Dot distance $r$. The half band-width $D$ sets the energy scale.
 
Correlations between the QD are the consequence of closed
electron-paths that involve both QD, paths driven by $H_V$ in
configurational space. The four lowest energy states are the
ferromagnetic triplet (FM, $|\!\! \uparrow\uparrow\rangle,
\ldots$; for which $\langle S_L.S_R \rangle = 1/4$ ) and the antiferromagnetic singlet (AF, $|\!
\downarrow\uparrow\rangle - |\! \uparrow\downarrow\rangle$; with $\langle S_L.S_R \rangle = -3/4$), with
an energy $-2E_d$. A well known second order (in $J_n$, four in
$\textbf{v}$) correlation is the RKKY process\cite{rkky, coq, cesar}, which splits the FM triplet from the AF state. Starting from the
FM state an electron (say the one in the left QD) hops to a band
state $k(>k_F)$ and then a second electron hops into the QD,
letting a hole $q(<k_F)$ in the band. The path is closed by
reversing these steps but at the other QD. The matrix element for
this loop is
\begin{eqnarray}\label{chsm}
\langle  \uparrow\uparrow |H_V^4|\!\uparrow \uparrow \rangle_{RKKY} =
 \mp \bf{v}^4 \cos{\bf{(k+q).r}}\ ,
\end{eqnarray}
where the plus sign holds for the AF state. This correlation path
produces an energy correction $\mp\Sigma_R(r)$ for the FM (AF)
state which splits the FM and AF states, it is easily evaluated by
perturbative\cite{cesar} or VWF methods\cite{jsfull}.

\begin{figure}[h]
\includegraphics[width=\columnwidth]{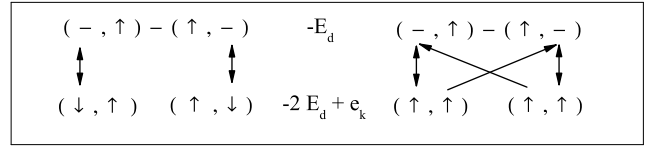}
\caption{First order correlation paths for the $|A_\uparrow
\rangle$ QD spin configuration. On the left it is shown the
non-correlated $\sigma \overline{\sigma}$  channel. On the right
it is the interference-enhanced $ \sigma \sigma $ channel, the
crossed arrows correspond to the inter-QD paths.}
\label{fig1}
\end{figure}

The proper one-electron-in-two-QD states are the symmetric and
antisymmetric localized spin $1/2$ states
\begin{equation}\label{as}
|A_\sigma \rangle =
\frac{1}{\sqrt{2}}(d^\dag_{R\sigma} \mp d^\dag_{L\sigma})|F\rangle=
\frac{1}{\sqrt{2}}(|0\sigma\rangle\mp|\sigma 0\rangle)\ ,
\end{equation}
where the plus sign holds for the symmetric state $|S_\sigma
\rangle$, these states have an energy $-E_d$.
They are the starting point of a first order correlation path. The
hybridization $H_V$ connect the $|A_\uparrow \rangle$ state, by
promoting an electron from below $k_F$ to the empty QD, with the
following ones
\begin{subequations}
\begin{eqnarray}\label{asmk}
|A_{\uparrow \downarrow k} \rangle = \frac{- \bf{v}}{\sqrt{2}} \
b^\dag_{k \uparrow} \ (e^{-i \bf{k.r}/2}\  |\!\uparrow \downarrow \rangle
+ e^{+i \bf{k.r}/2}\  |\!\downarrow \uparrow \rangle )\ , \\
\label{assk}
|A_{\uparrow \uparrow k} \rangle = \frac{- \bf{v}}{\sqrt{2}} \
b^\dag_{k \downarrow} \ (e^{-i \bf{k.r}/2}\  |\!\uparrow \uparrow \rangle
+ e^{+i \bf{k.r}/2}\ |\!\uparrow \uparrow \rangle )\ ,
\end{eqnarray}
\end{subequations}
of energy $-2E_d+e_k$. Closing the loops, as depicted in Fig.\ref{fig1},
the following matrix elements are obtained
\begin{subequations}
\begin{eqnarray}\label{chsma}
\langle A_\uparrow|H_V|A_{\uparrow \downarrow k} \rangle &=& \bf{v}^2\ , \\
\label{chsmb}
\langle A_\uparrow|H_V|A_{\uparrow \uparrow k} \rangle &=& \bf{v}^2(1\pm \cos{\bf{k.r}})\ ,
\end{eqnarray}
\end{subequations}
where the minus sign corresponds to a similar path but for the
$|S_\uparrow \rangle$ state. The last element depends on the
inter-QD distance $r$, thus it gives a correlation between the status
of the two QD.

As the starting state of the loop ($|A_\uparrow \rangle$)
has a higher energy than the bunches of ``visited" ones
($|A_{\uparrow\sigma k} \rangle$) the energy associated with this
correlation can not be obtained by perturbative methods
\cite{varma,lucio}. This situation, a single state connected to  bunches
of lower energy states, is the hallmark of a Kondo structure, in
this case a Kondo odd doublet that can be analyzed by means of the
following VWF\cite{jsfull, jsdoublet}
\begin{eqnarray}\label{wodd}
|D_{o\uparrow}\rangle= |A_\uparrow\rangle + \sum_{k}\  Z(k)
(|A_{\uparrow\downarrow k}\rangle+|A_{\uparrow\uparrow k}\rangle)\ .
\end{eqnarray}
Proceeding as in the one impurity case\cite{varma}, the
variational amplitude turns out to be $Z(k)=1/(E_o + 2E_d -e_k)$,
where $E_o$ is the energy of the doublet
\begin{eqnarray}\label{enedo}
E_o=-E_d+\textbf{v}^2\sum_{k}\frac{2 + \cos{k_x r}} {E_o + 2 E_d -
e_k }\ ,
\end{eqnarray}
assuming  $E_o = - 2 E_d - \delta_o$ (thus $Z(k)=-1/(\delta_o+
e_k)$) we obtain
\begin{equation}\label{es02}
E_d+\delta_o=\textbf{v}^2\sum_{k}\frac{2+\cos{k_x r}} {\delta_o+e_k}\ ,
\end{equation}
a self-consistent equation for the inter-QD correlation energy
$\delta_o(r)$. Two limits are easily found. For very large $r$ the
contribution from the $\cos$ term vanishes and thus $\delta_o$
tends to $\delta_K = D \exp{(-1/2J_n)}$, the one-impurity Kondo
energy. At $r=0$ ($\cos{k_x r}=1$) one obtains $\delta_o(0)=\delta_3=D
\exp{(-1/3J_n)}\gg \delta_K$\cite{lucio}. This is a huge correlation energy,
for $J_n=0.0723$ (i.e. $\delta_K=0.001 D$) one obtains a tenfold
increment over the one-QD  Kondo case, given that $\delta_3=0.01 D$.

For arbitrary inter-QD distances one must evaluate the sums in
Eq.(\ref{es02}). One is the Kondo Integral $I_K(\delta)= \Sigma_k
(-Z(k)/N) =  n_o \ln{(1+D/\delta)}$ and we call the other the
Quantum Coherence Integral $I_Q(\delta, r) = \Sigma_k (-\cos{(k_x
r)} Z(k)/N)$ which for a 1D LDEG becomes $I_Q(\delta, r) =
n_o(\cos{(y)} [ \text{Ci}(y)-\text{Ci}(x)] + \sin{(y)}[
\text{Si}(y)-\text{Si}(x)])\ $, where  $x=k_F r$ and
$y=x(1+\delta/D)$, $\text{Ci}$ ($\text{Si}$) is the CosIntegral
(SinIntegral) function\cite{jsdoublet}. The Coherence Integral has a logarithmic
dependence on $\delta$ that goes like $I_K$, note that
$I_Q(\delta,0) = I_K(\delta)$, so it is useful to define the
coherence factor
\begin{equation}\label{cqd}
C_Q(\delta,r)=I_Q(\delta,r)/I_K(\delta)\ ,
\end{equation}
which is a decaying oscillatory function that depends weakly on
$\delta$. For low dimensional or spatially confined LDEG, such
that the main decoherence factor is the energy width of the packet
of electrons involved in the interaction, $C_Q$ remains a relevant
factor up to inter-QD distances of the order of the Kondo screening length
($\xi_K = \lambda_F D/\delta$). The RKKY interaction, instead,
decays at a typical distance of the order of the Fermi length
$\lambda_F$. As a function of $C_Q$  Eq.(\ref{es02}) becomes
\begin{equation}\label{es04}
\delta_o(r)=D\ \exp{ \frac{-1}{[2 \pm C_Q(\delta_o, r)]\ J_n}
}\ ,
\end{equation}
that iteratively converges to the odd Doublet correlation energy
gain. The minus sign in front of $C_Q$ in Eq.(\ref{es04})
corresponds to the even Doublet energy $\delta_e(r)$, the even
Doublet is the Kondo structure that develops from
$|S_\sigma\rangle$, see Eqs.(\ref{wodd}),(\ref{chsmb}). Note that
$C_Q$ couples to the ``connectivity" factor, the $2$ in the
exponent of $\delta_K$, small changes in this factor produce
exponential changes in $\delta_o$.
\begin{figure}[h]
\includegraphics[width=\columnwidth]{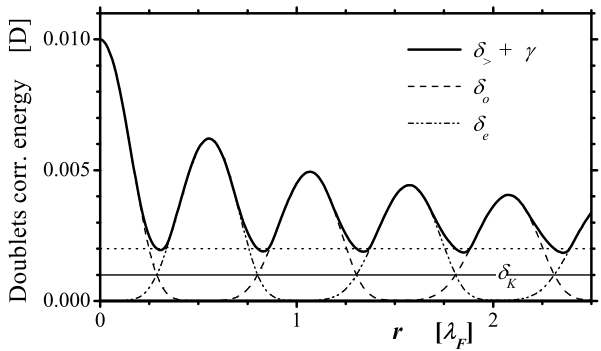}
\caption{Correlation energy gains $\delta_e$ and $\delta_o$ as a
function of $r$, for $J_n=0.07238$ ($\delta_K=0.001D$). These
energy gains are greater than $2\delta_K$ up to  $r\simeq\xi_K$.}
\label{fig2}
\end{figure}

In Fig.\ref{fig2} we plot $\delta_e$ and $\delta_o$ as function of
$r$. The periodicity with which $\delta_o$ and $\delta_e$
alternate ($\sim \lambda_F$) is determined by $C_Q$, $\delta_o$ is
greater than $\delta_e$ for $C_Q(r)>0$ and viceversa. They cross
each other at the $r_K$ points for which $C_Q(\delta_K,r_K)=0$,
and at these points both are equal to $\delta_K$.  For most
of the ($J_n, r $) range of interest these energies are greater than the RKKY energy $\Sigma_R$. In turn, these Doublets are connected through the Fermi Sea  state to form a composite Kondo Singlet with a further correlation energy gain $\gamma\leq\delta_K$ which is also plotted in Fig.\ref{fig2}. This second stage Kondo energy has its maxima at the zeros $C_Q$, at which $\gamma = \delta_K$, given for these points a total Kondo energy gain ($\delta+\gamma$) equal to $2\delta_K$. Besides these regions $\gamma$ is exponentially small. This composite singlet is mainly formed using the lower energy Doublet, which thus determines the QD-spin correlation. In Table \ref{table1} we list the main characteristics of the Kondo structures present in the two-QD Anderson-Kondo system.

\begin{table}
\caption{Vertex, bunch states (channels), connectivity factors (per channel
and total) and total energy for the Kondo structures in
the two-QD system. From top to bottom: one-impurity singlet, odd and even two-impurity Doublets and two-impurity composite Super-Singlet.} \label{table1}
\begin{ruledtabular}
\begin{tabular}{lccccc}
  & vertex & channel& chn.cnc. & tot.cnc. & energy\\
\hline
1I-S & $|F\rangle$ &  $\sigma $ & $1$  & $2$ & $ -E_d-\delta_K $ \\
 &&&&& \\
2I-$D_o$ & $|A_\sigma\rangle$ &  $\sigma \ \overline{\sigma} $ &  $1$  &  & \\
 &  & $\sigma \ \sigma$ &  $1 + C_q$  & $2 + C_q$ & $- 2 E_d - \delta_o$ \\
  &&&&& \\
2I-$D_e$ & $|S_\sigma\rangle$ &  $\sigma \ \overline{\sigma} $ &  $1$  &  & \\
 &  & $\sigma \ \sigma$ &  $1 - C_q$  & $2 - C_q$ & $- 2 E_d - \delta_e$ \\
 &&&&& \\
2I-SS & $|F\rangle$ &  $|D_{o\sigma}\rangle$\footnote{The channel
corresponding to the Doublet of lower energy (higher $\delta$) dominates
the Super-Singlet structure.} &  $1-C_q$  & $1-C_q$ & \\
 &  &  $|D_{e\sigma}\rangle$\footnotemark[1] &  $1+C_q$  & $1+C_q$
& $- 2 E_d - \delta_> -\gamma $\\
\end{tabular}
\end{ruledtabular}
\end{table}

For these Doublets the QD spin-spin correlation $\langle S_L.S_R\rangle$ is determined by
\begin{equation}\label{slsr}
\frac{\langle D| S_L.S_R |D\rangle}{\langle D|D\rangle}=
\pm \frac{3}{4}\ \frac{D_Q(\delta,r)}{(2\pm D_Q(\delta,r))} \ ,
\end{equation}
where $D_Q = (\partial_\delta I_Q)/(\partial_\delta I_K)$ and the
upper (lower) sign holds for the odd (even) Doublet.
\begin{figure}[h]
\includegraphics[width=\columnwidth]{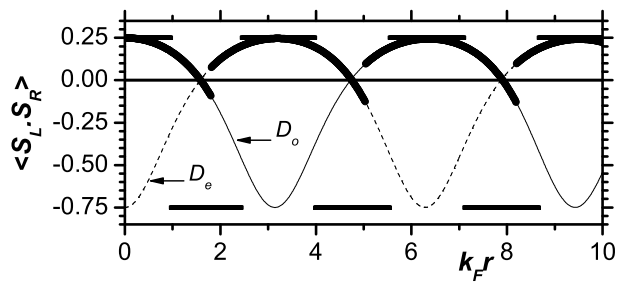}
\caption{QD-spin correlation for the doublets. The thicker section
in each line is the $r$-region dominated by the corresponding
Doublet. The straight segments at $1/4$ ($-3/4$) corresponds to
the FM (AF) state. } \label{fig3}
\end{figure}

In Fig.\ref{fig3} it can be seen that the Doublets favor a
ferromagnetic alignment of the QD spins. This is due to the fact that the enhanced channel is the $\sigma\sigma$ one, a ferro configuration (Eqs.(\ref{assk},\ref{chsmb})) and that the $\sigma\overline{\sigma}$ channel, which is a ferro or antiferro configuration depending on the relative phase of their components, is also mainly a ferro configuration at the maxima of $\delta$. Clearly, the screening action of the hole is more effective if the QD-spins are aligned and the distance between them is near the resonant condition for the lowest energy holes ($r\simeq n \lambda_F/2, |\cos{k_F r}|\simeq1$), see Eq.(\ref{chsmb}).   At the transition points from one Doublet to the other it is shown in Fig.\ref{fig3} a little negative value for $\langle S_L.S_R \rangle$ and then a jump to positive values. Actually, at those transition points the Super Singlet has a maximum in its correlation energy ($\gamma\simeq\delta_K$) and it is formed with similar weights in both Doublets, thus an average of the response of the odd and even Doublets  is to expect at those regions, given a smooth transition with a $\langle S_L.S_R \rangle$ value near zero from one Doublet region to the next one. Note that this mainly ferromagnetic QD-QD response is obtained without include any possible effect of the RKKY interaction in the Doublets structure. 

In Fig.\ref{fig3} it is shown also the RKKY prediction, based upon $\Sigma_R(r)\gtrless 0$ ($\Sigma_R(r)$ is equal to $4 \ln{2}\ J_n^2\ (1-2 \text{Si}(2k_F r)/\pi)\ D$ in 1D\cite{litvi}). Taken into account both interactions, the stronger of them determines the $\langle S_L.S_R \rangle$ response of the system, for $\max{(\delta_o,\delta_e)} \gg |\Sigma_R|$ the response is determined by the Doublets curve (Eq.(\ref{slsr}), Fig.\ref{fig3}) whereas that in the opposite situation the QD-spin correlation tends to that of the RKKY.

\begin{figure}[h]
\includegraphics[width=\columnwidth]{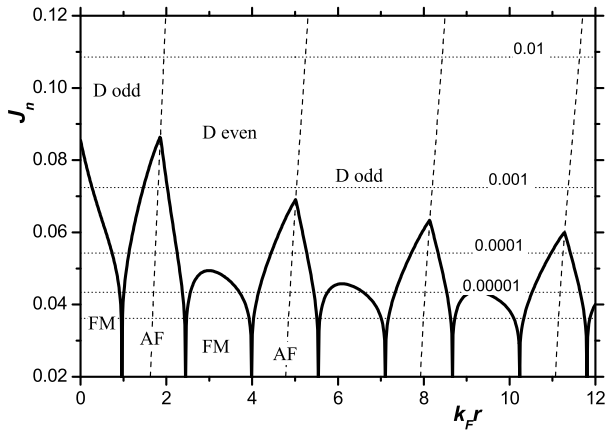}
\caption{$\langle S_L.S_R \rangle$ 1D ``Quantum Phase Diagram" for the two-QD spin correlation. The horizontal dot lines mark the values of $\delta_K$ (in units of $D$) for the corresponding values of $J_n$.} \label{fig4}
\end{figure}

In Fig.\ref{fig4} we show the $\langle S_L.S_R \rangle$ ``Parameters Phase Diagram" for the 1D two-QD system, determined by $\max{(\delta_o,\delta_e)}=|\Sigma_R|$. This is not a true Quantum Phase diagram given that not the Doublets nor the FM (AF) states are the ground state of the system, it indicates the tendency of the QD-spin correlation to follow a Doublet-like or RKKY-like behavior. Over the near vertical dashed lines that mark the limits between the odd and even Doublets, determined by $C_Q(\delta_K)=0$, there is little enhancement of the Doublets energies. These zones coincide with the AF maxima of the RKKY interaction and thus over these lines there are the highest indents of the RKKY-AF states, which favor an antiferromagnetic alignment of the QD spins. Note also that in the RKKY-FM regions (i.e. $\Sigma_R \gg \delta_>$) the effect of the corresponding Doublet is to screen the total Spin of the system\cite{jones} but, simultaneously, it reinforces the ferromagnetic QD-spin response. 

In this work we have shown how a first-order correlation present in the Two-Impurity Anderson Hamiltonian leads to the formation of a pair of Kondo Doublet states. The localized spins correlation $\langle S_L.S_R \rangle$ response for these Doublets strongly differs from the one predicted by the second-order RKKY interaction (Fig.\ref{fig3}). These Kondo Doublets generates a mainly ferromagnetic QD-spin correlation without resort to the second-order RKKY interaction.   The correlation energy gain of these Doublets, which is driven by direct coherence effects of the hybridization, exceeds that of the RKKY interaction in most of the Hamiltonian parameter-space ($(J_n, r)$, Fig.\ref{fig4}). A remarkable property of this interaction is the long distances ($\sim \xi_K$) at which the impurities remain in a correlated state. This is due to the fact that just the electrons whose energy differs from that of the Fermi level less than $\delta$  contribute significatively to this correlation. Angular decoherence effects reduce this range for a higher-Dimension electron bath gas. These properties put these Doublets states well in the experimentally accessible range for QD  systems built on  semiconductor devices\cite{qddevices}, were the most relevant parameters can be controlled
by gate voltages. The VWF method we use can be extended to analyze systems with several QD.

 I am a fellow of the CONICET
(Argentina), which partially financed this research under grant
PIP 02753/00.

\bibliography{kondo}

\end{document}